%% file: main.tex
\documentclass[journal,trans]{IEEEtran}
\usepackage{hyperref,color,breqn,multirow}
\usepackage[top=0.7in, left=0.65in]{geometry}
\usepackage{color}
\usepackage{graphicx}
\usepackage{amsmath} 
\usepackage{amssymb}
\usepackage{amsthm,bm}
\usepackage{siunitx}
\usepackage{eso-pic}

\AddToShipoutPicture*{\footnotesize\sffamily\raisebox{1cm}{\hspace{1.65cm}\fbox{\parbox{\textwidth}{\copyright 2023 IEEE.  Personal use of this material is permitted.  Permission from IEEE must be obtained for all other uses, in any current or future media, including reprinting/republishing this material for advertising or promotional purposes, creating new collective works, for resale or redistribution to servers or lists, or reuse of any copyrighted component of this work in other works.}}}}

\usepackage{tikz}
\usepackage{pgfplots}
\pgfplotsset{width=10cm,compat=1.9}
\usepgfplotslibrary{fillbetween}

\begin{document}

\title{Combination of Measurement Data and Domain Knowledge for Simulation of Halbach Arrays with Bayesian Inference}

\author{\IEEEauthorblockN{Luisa Fleig\IEEEauthorrefmark{1,2},
			Melvin Liebsch\IEEEauthorrefmark{1},
			Stephan Russenschuck\IEEEauthorrefmark{1}, and 
            Sebastian Schöps\IEEEauthorrefmark{2}}
\IEEEauthorblockA{\IEEEauthorrefmark{1}European Organization for Nuclear Research (CERN), Meyrin, Switzerland}
\IEEEauthorblockA{\IEEEauthorrefmark{2}Technical University of Darmstadt, Darmstadt, Germany}
\thanks{ (email: luisa.fabiola.fleig@cern.ch).}}

\markboth{Journal of \LaTeX\ Class Files,~Vol.~14, No.~8, August~2015}%
{Shell \MakeLowercase{\textit{et al.}}: Bare Demo of IEEEtran.cls for IEEE Transactions on Magnetics Journals}

\IEEEtitleabstractindextext{%
\begin{abstract}

            Accelerator magnets made from blocks of
			permanent magnets in a zero-clearance configuration are known as Halbach arrays. 
			The objective of this work is the fusion of knowledge from different measurement sources (material and field) and domain knowledge (magnetostatics) to obtain an updated magnet model of a Halbach array.
			From Helmholtz-coil measurements of the magnetized blocks, a prior distribution of the magnetization is estimated. Measurements of the magnetic flux density are used to derive, by means of Bayesian inference, a posterior distribution.
            The method is validated on simulated data and applied to measurements of a dipole of the FASER detector. The updated magnet model of the FASER dipole describes the magnetic flux density one order of magnitude better than the prior magnet model.

\end{abstract}

\begin{IEEEkeywords}
Bayesian inference, Inverse problems, Permanent magnets, Hallbach array, Monte Carlo methods
\end{IEEEkeywords}}

\maketitle

\IEEEdisplaynontitleabstractindextext

\IEEEpeerreviewmaketitle

\section{Introduction}

\begin{figure}
	\centering
    \includegraphics[width=0.5\linewidth]{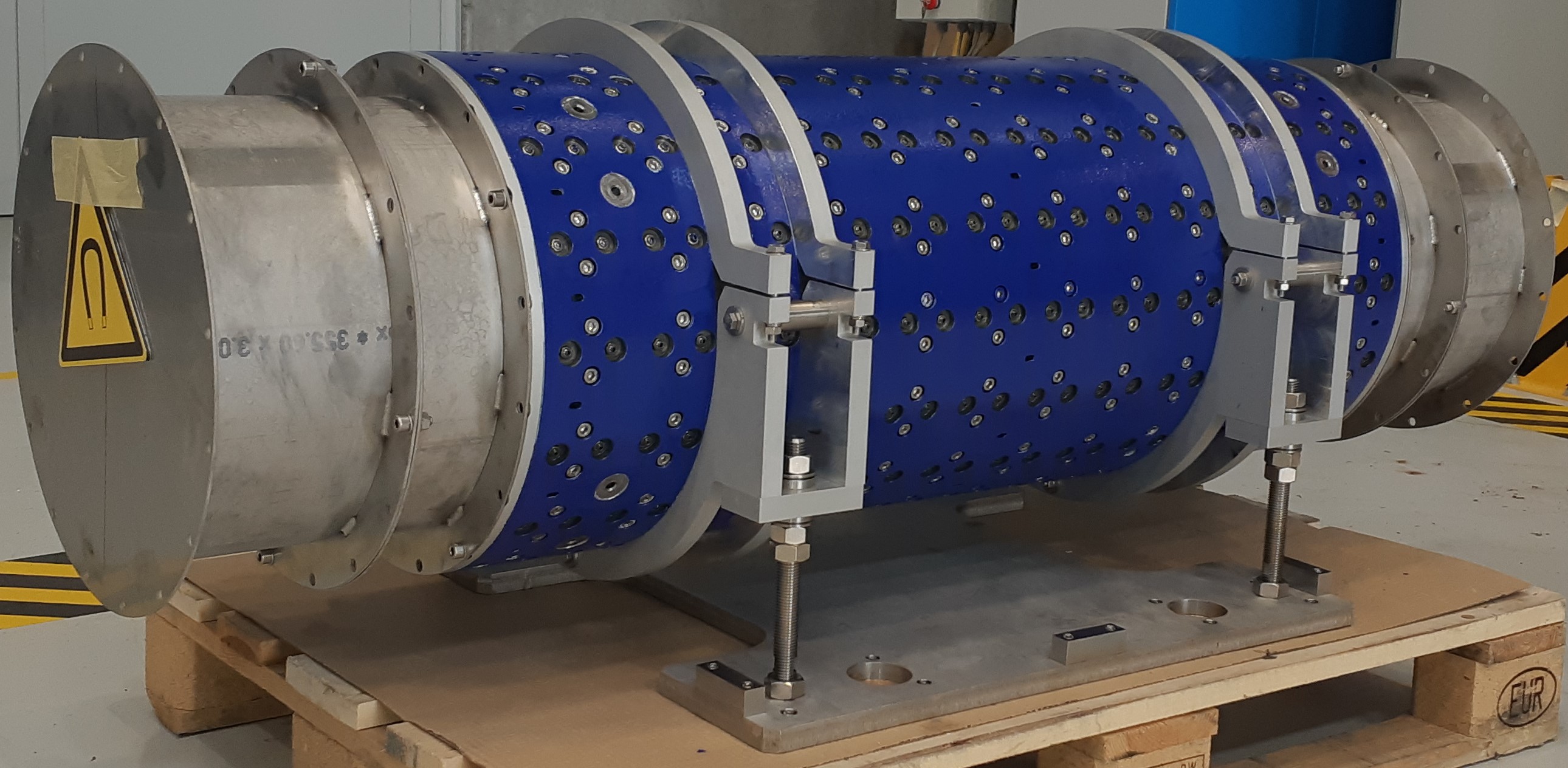}\\[-0.5em]
	\caption{FASER dipole magnet.}
	\label{fig:FASERdipole}
	\vspace{-1em}
\end{figure}

\IEEEPARstart{A}{}special type of accelerator magnet, referred to as Halbach array \cite{Halbach}, is made of circularly arranged blocks of permanent magnets (PM). The magnet system of the FASER detector \cite{Faser} consists of three \SI{0.57}{\tesla} Halbach dipole magnets. In \cite{messreport}, it is shown that imperfections of the PM block magnetizations affect the field quality by generating higher-order multipole field errors. 
\newline Quality assurance of the FASER dipole, see \autoref{fig:FASERdipole}, consisted of two measurement campaigns \cite{messreport}: (i) The measurement of the magnetization of each PM block before the magnet assembly using a Helmholtz-coil system and (ii) the measurement of the magnetic flux density of the built magnet with a 3D Hall probe mapper. In both cases deviations arise due to sensor calibration, stage alignment, magnetization errors in the PM blocks and manufacturing errors in the magnet assembly.

Deriving the magnetization of PM blocks from measured magnetic flux densities  is a well known inverse problem \cite{Kovacs, Arbenz, ChavinCollin}, in particular for Halbach arrays \cite{Chavin, Bruckner}. Different methods and regularization techniques are used, e.g., physics informed neural networks \cite{Kovacs}, singular value decomposition \cite{Chavin, Arbenz}, Tikhonov regularization \cite{Bruckner} and Bayesian inference based on a linear model \cite{ChavinCollin}.
\newline In this work, Bayesian inference is used to combine the prior magnetization data obtained from the Helmholtz-coil measurements with the magnetic flux density measurements in the magnet, taken with the Hall probe mapper and the domain knowledge imposed by the magnetostatic problem. 
This method is validated on simulated data and applied to measured data of a FASER dipole. Inserting the posterior magnetization in the simulation leads to an updated model that predicts the magnetic flux density in the homogeneous field region one order of magnitude better than the prior simulation model.

\section{Problem statement}

\begin{figure}
	\begin{minipage}[t]{0.25\textwidth}
    \def\svgwidth{0.80\linewidth}
    \tiny
    \input{Halbach2D.txt}
    \end{minipage}%
    \hspace*{-2em}
    \begin{minipage}[t]{0.25\textwidth}
	\def\svgwidth{0.91\linewidth}
    \normalsize
    \input{Halbach_log.txt}
    \end{minipage}
    \caption{
        Left: Domain $D$ with the cross section of the FASER Halbach dipole with PM blocks $D_i$ and their nominal magnetization $\overline{\mathbf{M}}_i$. Right: Absolute value of the Gateaux derivative $\mathbf{B}'$ of the mapping \eqref{eq:mapping13}.
	\label{fig:Halbach+Gateaux}
    }
\end{figure}
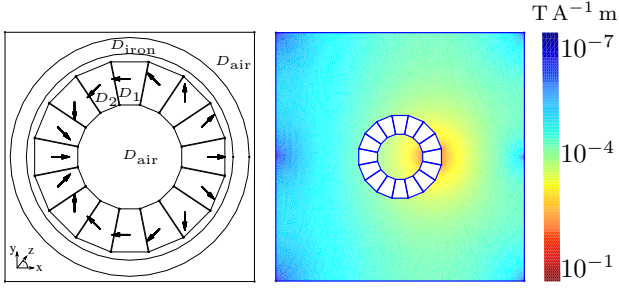

The magnetostatic problem in a Halbach array on a domain $D=D_{\mathrm{air}}\cup D_{\mathrm{m}}\cup D_{\mathrm{iron}}$  is defined by $ \mathrm{curl}\: \mathbf{H}=\mathbf{0}$ in $D$, $\mathrm{div}\:\mathbf{B}=0 $ in $ D$, $\mathbf{B}\cdot \mathbf{n}=0 $ on $\partial D, $
with magnetic flux density $\mathbf{B}$, magnetic field strength $\mathbf{H}$ and outward pointing unit normal vector $\mathbf{n}$. The corresponding constitutive equation is $\nu(\left\|\mathbf{B} \right\| ) \mathbf{B} = \mathbf{H} + \mathbf{M}$, 
where $\mathbf{M}$ denotes the magnetization with $\mathrm{supp}(\mathbf{M})=D_{\mathrm{m}}=\bigcup_{i=1}^{16} D_i$ and $D_i$ denote the PM blocks (\autoref{fig:Halbach+Gateaux} left).
 The reluctivity function $\nu:\mathbb{R}^+_0\rightarrow\mathbb{R}^+_0$ is defined by
\begin{dmath}
	\nu(s) =	\begin{cases} 1/\mu_0 & \mathrm{in}\;D_{\mathrm{air}}\\
			1/(\mu_0 \mu_r) & \mathrm{in}\;D_{\mathrm{m}}\\
			f_{\mathrm{HB}}(s)/s & \mathrm{in}\;D_{\mathrm{iron}}		
	\end{cases}		
\end{dmath}
dependent on a (non-linear) $H(B)$-curve $f_{\mathrm{HB}}$ on the iron domain $D_{\mathrm{iron}}$.

Insertion of the magnetic vector potential $\mathbf{A}$ defined by $\mathrm{curl}\;\mathbf{A}=\mathbf{B}$ and the constitutive equation in the magnetostatic problem yields the well-known curl-curl formulation of the problem which is discretized and solved by the finite element (FE) software getDP \cite{Geuzaine}.

The FASER dipoles are composed of a series of 12 to 18 rings with 16 PM blocks per ring. We use the index $i$ to indicate the PM block number in a ring and index $j$ to indicate the ring number. The nominal magnetization of the PM blocks $(D_i)_{i=1,\dots,16}$ is parametrized by
\begin{equation}
    \label{M_nom}
    \overline{\mathbf{M}}_{i,j}
    :=
    \overline{m}
    \;
    \textrm{vol}(D_i)^{-1}   
    \;
    (
        \cos(\overline{\alpha}_i),
        \sin(\overline{\alpha}_i),
        0
    )^\top
\end{equation}
for all $j=1,\dots,12$, with nominal orientation $\overline{\alpha}_i:=180\unit{\degree}+2(i-1)\cdot 22.5\unit{\degree}$ (\autoref{fig:Halbach+Gateaux} left) and nominal magnetic moments $\overline{m}_i=\SI{330}{\A\square\m}$. 
Due to manufacturing tolerances and production errors, the magnetizations $\mathbf{M}_{ij}$ of the procured PM blocks differ from the nominal (specified) values. This can be expressed as
\begin{equation}
    \mathbf{M}_{i,j}
    =
    \overline{\mathbf{M}}_{i,j}
    +
    {\Delta\mathbf{M}}_{i,j},
\end{equation}
adding the magnetization deviation ${\Delta\mathbf{M}}_{i,j}$. The magnetization of each PM block was measured by means of a Helmholtz coil prior to the assembly of the Halbach array \cite{messreport}. We interpret ${\Delta\mathbf{M}}_{i,j}$ and $\mathbf{M}_{i,j}$ as random vectors that can be deducted from the measurements.

We define the parameter vector $\mathbf{p}$ of the PM block magnetization by
\begin{equation}
    \mathbf{p}:=(\mathbf{M}^x_{i,j},\mathbf{M}^y_{i,j},\mathbf{M}^z_{i,j}):\Omega\to\mathbb{R}^{16\times 12\times 3}
\end{equation}
in the 3D case while the 2D case is analogous for only the $x$ and $y$ components;

$\Omega$ is the probability space that reflects the random occurrence of production errors of the PM blocks. Having measurements for multiple PM blocks of the same magnetization type, corresponding to different rings of the Halbach dipole, a prior distribution $\pi_0$ of $\mathbf{p}$ can be derived by computing the sample mean $\bm{\mu}$ and sample covariance $\mathbf{C}_0$, leading to $\mathbf{p}\sim \pi_0:=\mathcal{N}(\bm{\mu}_0,\mathbf{C}_0)$, where $\mathcal{N}$ is the Gaussian distribution. The Anderson-Darling test justifies the usage of the Gaussian distribution.

The simulation model
    $\mathcal{H}: \mathbf{p}\mapsto \mathbf{q}$
maps the parameter vector $\mathbf{p}$ to a field-related measurable quantity $\mathbf{q}$ such as the magnetic flux density $\mathbf{B}$ in the bore of the magnet. 
The two definitions of $\mathbf{q}$ that are used in this paper are
\begin{equation}
\mathbf{q}_{\mathbf{B}}:=\left(
\mathbf{B}_x
,
\mathbf{B}_y
,
\mathbf{B}_z
\right),
\end{equation}
where each $\mathbf{B}_\star$ depends on measurement positions $\mathbf{x},\mathbf{y},\mathbf{z}$, and
\begin{equation}
\mathbf{q}_{\mathrm{F}}:=\left(
    A_1
    ,
    \dots,
    A_K
    ,
    B_1
    ,
    \dots
    ,
    B_K
\right),
\end{equation}
where $A_k(\mathbf{z})$ and $B_k(\mathbf{z})$ are the Fourier coefficients
of the magnetic flux density component $\mathbf{B}_r(\mathbf{z})$ on a circle with radius $r_0=\SI{75}{\milli\meter}$, centered in the magnet bore. The advantage of working with the Fourier coefficients is the averaging effect to random noise. However, especially in the 2D case, considering only the Fourier coefficient leads to an under-determined inverse problem.

If the reluctivity in $D$ is assumed to be constant, e.g. the iron ring is neglected $D_{\mathrm{iron}}=\emptyset$, the  model is linear and we can write $\mathcal{H}\mathbf{p}=\mathbf{q}$ by abuse of notation.

To select positions $\mathbf{x},\mathbf{y}$, for which the magnetic flux density $\mathbf{B}$ is sensitive to variations of $\mathbf{p}$, a sensitivity analysis is performed. Similar to \cite{Roemer} it can be shown that the Gateaux derivative $\mathbf{B}'$ of the mapping 
\begin{equation}
    {\Delta\mathbf{M}} \mapsto \mathbf{B}[\overline{\mathbf{M}}+{\Delta\mathbf{M}}]
\end{equation}
can be determined by finding the weak solution $\mathbf{A}'$ of
\begin{equation*}
\int_{D}\!\nu(|\mathrm{curl}\:\mathbf{A}'|)\mathrm{curl}\:\mathbf{A}'\cdot \mathrm{curl}\:\mathbf{v}\:\mathrm{d}V
=\!\!\int_{D}\!\mathrm{curl}\:{\Delta\mathbf{M}}\cdot \mathrm{curl}\:\mathbf{v}\:\mathrm{d}V
\end{equation*}
and computing $\mathbf{B}'=\mathrm{curl}\:\mathbf{A}'$. For illustration, \autoref{fig:Halbach+Gateaux} shows the absolute value of the Gateaux derivative
\begin{equation}
    \label{eq:mapping13}
    \mathbb{E}[{\Delta\mathbf{M}}_{13}]\mapsto \mathbf{B}\left[\overline{\mathbf{M}}_{13}+\mathbb{E}[{\Delta\mathbf{M}}_{13}]\right].
\end{equation}
It increases towards the segment whose magnetization is varied. Thus, equally distributed measurement positions on a centered circle inside the inner air region of the magnet are chosen.

\section{Bayesian inference}
Due to uncertainties of the magnetic flux density measurements, the measurable quantity $\mathbf{q}=\mathcal{H}(\mathbf{p})$ is not directly observable, but $\mathbf{q}^{\mathrm{obs}}:=\mathcal{H}(\mathbf{p})+\bm{\varepsilon}$ with $\bm{\varepsilon}\sim \mathcal{N}(\mathbf{0},\bm\Sigma)$. Thus, the distribution of $\mathbf{q}^{\mathrm{obs}}$ given $\mathbf{p}$ is $\pi(\mathbf{q}^{\mathrm{obs}}|\mathbf{p})=\mathcal{N}(\mathcal{H}(\mathbf{p}),\bm\Sigma)$.
	Following \cite{Bardsley}, Bayes' rule 
	\begin{dmath}\pi(\mathbf{p}|\mathbf{q}^{\mathrm{obs}})=\frac{\pi(\mathbf{q}^{\mathrm{obs}}|\mathbf{p})\pi_0(\mathbf{p})}{\pi(\mathbf{q}^{\mathrm{obs}})}
	\end{dmath}
	can be applied to determine the posterior distribution of the magnetization parameter vector $\mathbf{p}$, given an observation $\mathbf{q}^{\mathrm{obs}}$ of the measurable quantity.
    To generate samples from the posterior distribution, we use different approaches depending on whether the model $\mathcal{H}$ is linear or nonlinear.
    
    In the \emph{linear case}, the posterior distribution is proportional to a Gaussian distribution \cite{Bishop}
    \begin{equation}
        \pi(\mathbf{p}|\mathbf{q}^{\mathrm{obs}})\propto \mathcal{N}(\bm\mu_1,\mathbf{C}_1).
    \end{equation}
    The expected value $\bm\mu_1$ and the covariance matrix $\mathbf{C}_1$ of this distribution are given by
    \begin{eqnarray}
        \bm\mu_1&=&\mathbf{C}_1\left( \mathcal{H}^\top\boldsymbol\Sigma^{-1}\mathbf{q}^{\mathrm{obs}}+\mathbf{C}^{-1}_0 \bm\mu_0\right), \\
        \mathbf{C}_1 &=&\left(\mathcal{H}^\top\boldsymbol\Sigma^{-1}\mathcal{H}+ \mathbf{C}^{-1}_0\right)^{-1}.
    \end{eqnarray}

    
    In the \emph{nonlinear case}, the Metropolis-Hastings algorithm \cite{Bardsley} is applied. The main idea is the construction of a Markov chain, whose states are samples of $\pi(\mathbf{p}|\mathbf{q}^{\mathrm{obs}})$. Therefore, the transition kernel $K$, for which $\pi(\mathbf{p}|\mathbf{q}^{\mathrm{obs}})$ is an invariant distribution, is defined by
	\begin{equation}
	    K(\hat{\mathbf{p}}|\mathbf{p}):=a(\hat{\mathbf{p}},\mathbf{p})\cdot\tilde{\pi}(\hat{\mathbf{p}}|\mathbf{p})
	\end{equation}
	with the proposal density $\tilde{\pi}(\hat{\mathbf{p}}|\mathbf{p})$ and
	\begin{equation}\label{def_a}
		a(\hat{\mathbf{p}},\mathbf{p}):=\min\left\lbrace 1, \frac{\pi(\hat{\mathbf{p}}|\mathbf{q}^{\mathrm{obs}})\tilde{\pi}(\mathbf{p}|\hat{\mathbf{p}})}{\pi(\mathbf{p}|\mathbf{q}^{\mathrm{obs}})\tilde{\pi}(\hat{\mathbf{p}}|\mathbf{p})}\right\rbrace.
	\end{equation}
	The preconditioned Crank Nicolson proposal \cite{Neal}
	\begin{equation}
		\tilde{\pi}(\hat{\mathbf{p}}|\mathbf{p}):=\mathcal{N}(\sqrt{1-s^2}\mathbf{p},s\mathbf{C}_0)
	\end{equation}
	with step size $s=1/80$ is chosen, which incorporates the prior knowledge on the covariance $\mathbf{C}_0$.
	The sample $\left\lbrace \mathbf{p}_k\right\rbrace _{k\geq 0}$ is then generated by conducting the following steps of the Metropolis-Hastings sampling \cite{Bardsley}:
	\begin{enumerate}
		\item Initialize $\mathbf{p}_0:=\boldsymbol\mu_0$. Set $k=1$.
		\item Generate proposed sample $\tilde{\mathbf{p}}\sim \tilde{\pi}(\tilde{\mathbf{p}}|\mathbf{p}_{k-1})$. Accept $\tilde{\mathbf{p}}$ by setting $\mathbf{p}_k:=\tilde{\mathbf{p}}$ with probability $a(\tilde{\mathbf{p}},\mathbf{p}_{k-1})$. Else, set $\mathbf{p}_k:=\mathbf{p}_{k-1}$.
		\item Set $k=k+1$. Return to step 2. 
	\end{enumerate}
	Inserting the Gauss distributions of the prior, the measurement uncertainty and the preconditioned Crank Nicolson proposal in \eqref{def_a} simplifies the acceptance probability to
	\begin{equation*}
		a(\tilde{\mathbf{p}},\mathbf{p}_{k-1})=\min\!\left\lbrace\!1, \frac{\exp\left( \frac{1}{2} \left\| \mathcal{H}(\mathbf{p}_{k-1})-\mathbf{q}^{\mathrm{obs}}\right\| ^2_{\Sigma^{-1}}\right) }{\exp\left( \frac{1}{2} \left\| \mathcal{H}(\tilde{\mathbf{p}})-\mathbf{q}^{\mathrm{obs}}\right\| ^2_{\Sigma^{-1}}\right) }\!\right\rbrace\!.
	\end{equation*}

\section{Validation}

For validation of the algorithm, observation data is generated based on a ground truth parameter vector $\mathbf{p}^{\mathrm{true}}\sim\pi_0$ by setting
\begin{equation}
    \mathbf{q}^{\mathrm{obs}}:=\mathcal{H}(\mathbf{p}^{\mathrm{true}})+\bm{\varepsilon}, \qquad \bm{\varepsilon}\sim \mathcal{N}(\mathbf{0},\sigma^2\mathbb{I}).
\end{equation}
For $\mathbf{q}^{\mathrm{obs}}_{\mathbf{B}}$ based on magnetic flux densities we set $\sigma=\SI{e-4}{\tesla}$, for the observation of Fourier coefficients $\mathbf{q}^{\mathrm{obs}}_{\mathrm{F}}$ we set $\sigma=\SI{e-6}{\tesla}$, which are reasonable values for modern day measurement equipment.

\begin{figure}
	\centering
    \def\svgwidth{\linewidth}
    \scriptsize
    \input{lin_validation.tex}
    \caption{Validation of posterior derivation algorithm in linear case on 3D simulation model. Comparison of ground truth (red), prior $\mathcal{N}(\bm\mu_0,\mathbf{C}_0)$ (black), posterior based on magnetic flux density observation $\mathcal{N}(\bm\mu^{\mathbf{B}}_1,\mathbf{C}^{\mathbf{B}}_1)$ (blue) and posterior based on magnetic Fourier coefficient observation $\mathcal{N}(\bm\mu^{\mathrm{F}}_1,\mathbf{C}^{\mathrm{F}}_1)$ (green).}
	\label{fig:lin_validation}
\end{figure}
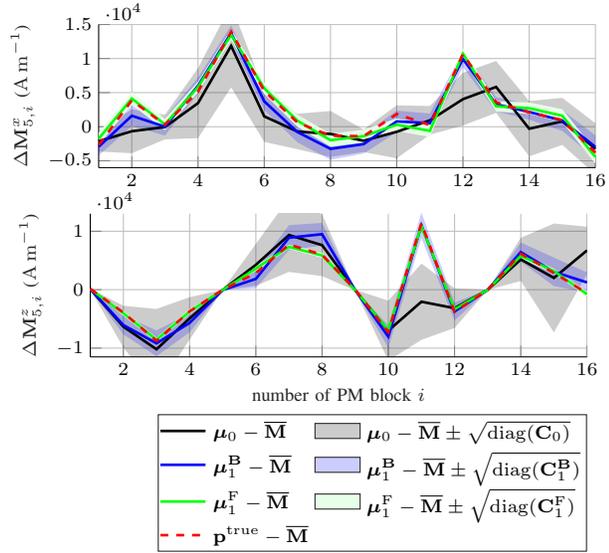

In the \emph{linear case}, the algorithms are validated on the 3D simulation model.
The ground truth $\mathbf{p}^{\mathrm{true}}$ is compared to the prior distribution $\pi_0$ and the two posterior distributions 
\begin{eqnarray}
\pi^{\mathbf{B}}(\mathbf{p}|\mathbf{q}^{\mathrm{obs}}_{\mathbf{B}})&=&\mathcal{N}(\bm\mu^{\mathbf{B}}_1,\mathbf{C}^{\mathbf{B}}_1)\\
\pi^{\mathrm{F}}(\mathbf{p}|\mathbf{q}^{\mathrm{obs}}_{\mathrm{F}})&=&\mathcal{N}(\bm\mu^{\mathrm{F}}_1,\mathbf{C}^{\mathrm{F}}_1)
\end{eqnarray}
that are based on the different observation vectors. In Figure ~\ref{fig:lin_validation}, the ground truth of the magnetization deviation $\mathbf{p}^{\mathrm{true}}-\overline{\mathbf{M}}$ restricted to the 5th ring of the Halbach array is shown together with the expected values and the variances. The maximal difference between $\bm\mu^{\mathbf{B}}_1$ and $\mathbf{p}^{\mathrm{true}}$ is decreased by 70\% compared to the maximal difference between $\bm\mu_0$ and $\mathbf{p}^{\mathrm{true}}$. The variances are also smaller.

\begin{figure}
	\centering
    \def\svgwidth{\linewidth}
    \scriptsize
    \input{update_MCMC.tex}
    \caption{Validation of posterior derivation algorithm in non linear case on 2D simulation model. Comparison of ground truth (red), prior $\mathcal{N}(\bm\mu_0,\mathbf{C}_0)$ (black) and posterior (blue) sample mean and variance.}
	\label{distributions}
\end{figure}
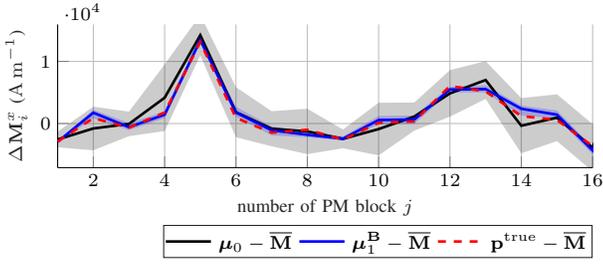

In the \emph{nonlinear case}, the method is applied to the 2D simulation model and simulated observations $\mathbf{q}^{\mathrm{obs}}_{\mathbf{B}}$ of magnetic flux densities.
To obtain the sample $\left\lbrace\mathbf{p}_k \right\rbrace_k $ the Metropolis-Hastings sampling is applied for $k=18000$ steps. In Figure~\ref{distributions} the sample mean and the sample variance of the prior and posterior distributions are plotted with the ground truth $\mathbf{p}^{\mathrm{true}}$. The posterior sample mean follows the ground truth better than the prior sample mean, the maximal difference is decreased by 50\%.

\section{Application}
The algorithm in the linear case (neglecting the outer iron ring) is applied to observations $\mathbf{q}^{\mathrm{obs}}_{\mathrm{F}}$ of $K=8$ Fourier coefficients on a centered circle with $r_0=\SI{75}{\milli\meter}$ radius in $\mathrm{dim}(\mathbf{z})=156$ positions along the full range of the magnet, including the fringe field. Measurements of the magnetic flux density are taken with a Hall probe mapper for each coordinate of $\mathbf{z}$ in $60$ equally distributed points on the circle and $\mathbf{q}^{\mathrm{obs}}_{\mathrm{F}}$ is obtained by Fourier analysis.

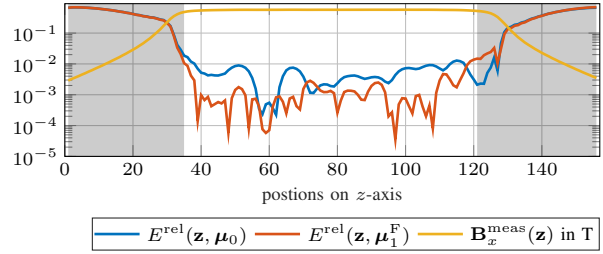
\begin{figure}
	\centering
    \def\svgwidth{1.5\linewidth}
    \scriptsize
    \input{update.tex}
    \caption{Relative error of prior and posterior simulation model compared to magnetic flux density measurements of the first FASER dipole. Area of fringe field marked in grey.}
	\label{fig:real}
\end{figure}

For estimating the noise covariance matrix $\bm\Sigma$ positioning uncertainties that result from
small misalignments between the various coordinate systems (magnet, mapper, simulation) as well as small oscillations of the Hall probe mapper system have to be taken into account. In the fringe field region, where the magnetic flux density changes the most, positioning errors lead to larger measurement errors than in regions where the magnetic flux density is almost constant. Therefore, $\Sigma=\sigma(z)\mathbb{I}$ is chosen position-dependent with
\begin{equation*}
    \sigma(z)=
    \begin{cases}
        \SI{5e-05}{\tesla} & z \textrm{ in the homogen. field region,}\\
        \SI{5e-03}{\tesla} & z \textrm{ in the fringe field}.
    \end{cases}
\end{equation*}
Figure~\ref{fig:real} shows the relative error
\begin{equation}
    E^{\mathrm{rel}}(\mathbf{z},\mathbf{p})=\left| \frac{\mathbf{B}^{\mathrm{meas}}_x(\mathbf{z}) - \mathbf{B}^{\mathrm{sim}}_x(\mathbf{z},\mathbf{p})}{\mathbf{B}^{\mathrm{meas}}_x(\mathbf{z})} \right|
\end{equation}
between the measured and simulated $\mathbf{B}_x$ component along the $z$-axis for the simulation with the prior parameter vector $\bm\mu_0$ and the updated posterior parameter vector $\bm\mu_1^{\mathrm{F}}$. The measured flux density data used in this comparison was not part of the training set used for updating the model.  The relative error $E^{\mathrm{rel}}(\mathbf{z},\bm\mu_1^{\mathrm{F}})$ of the update is one order of magnitude smaller than the relative error $E^{\mathrm{rel}}(\mathbf{z},\bm\mu_0)$ of the prior, almost everywhere in the homogeneous field region.

\section{Conclusion}

Bayesian inference was applied to combine domain knowledge and obervations from material and magnetic flux density measurements for deriving a posterior magnetization distribution of the PM blocks of an Halbach array. The method is not only validated on simulated data, but also applied to measurements of a Halbach dipole of the FASER experiment. Updating the magnetization of the PM blocks with the expected value of the posterior distribution decreases the relative error of the simulated magnetic flux density compared to the measured magnetic flux density inside the aperture by one order of magnitude. A more detailed analysis of the systematic measurement errors will be investigated in future research.

\section*{Acknowledgment}
This work has been supported by the Gentner Programme of the German Federal Ministry of Education and Research, and the Grad. School for Comp. Eng. at TU Darmstadt. The authors like to thank Erik Schnaubelt for helping with the getDP models and Olaf Dunkel and Mariano Pentella for conducting the PM block measurements.

\end{document}

%% file: Halbach2D.txt
\begingroup%
  \makeatletter%
  \providecommand\color[2][]{%
    \errmessage{(Inkscape) Color is used for the text in Inkscape, but the package 'color.sty' is not loaded}%
    \renewcommand\color[2][]{}%
  }%
  \providecommand\transparent[1]{%
    \errmessage{(Inkscape) Transparency is used (non-zero) for the text in Inkscape, but the package 'transparent.sty' is not loaded}%
    \renewcommand\transparent[1]{}%
  }%
  \providecommand\rotatebox[2]{#2}%
  \newcommand*\fsize{\dimexpr\f@size pt\relax}%
  \newcommand*\lineheight[1]{\fontsize{\fsize}{#1\fsize}\selectfont}%
  \ifx\svgwidth\undefined%
    \setlength{\unitlength}{848.21118164bp}%
    \ifx\svgscale\undefined%
      \relax%
    \else%
      \setlength{\unitlength}{\unitlength * \real{\svgscale}}%
    \fi%
  \else%
    \setlength{\unitlength}{\svgwidth}%
  \fi%
  \global\let\svgwidth\undefined%
  \global\let\svgscale\undefined%
  \makeatother%
  \begin{picture}(1,0.99493671)%
    \lineheight{1}%
    \setlength\tabcolsep{0pt}%
    \put(0,0){\includegraphics[width=\unitlength]{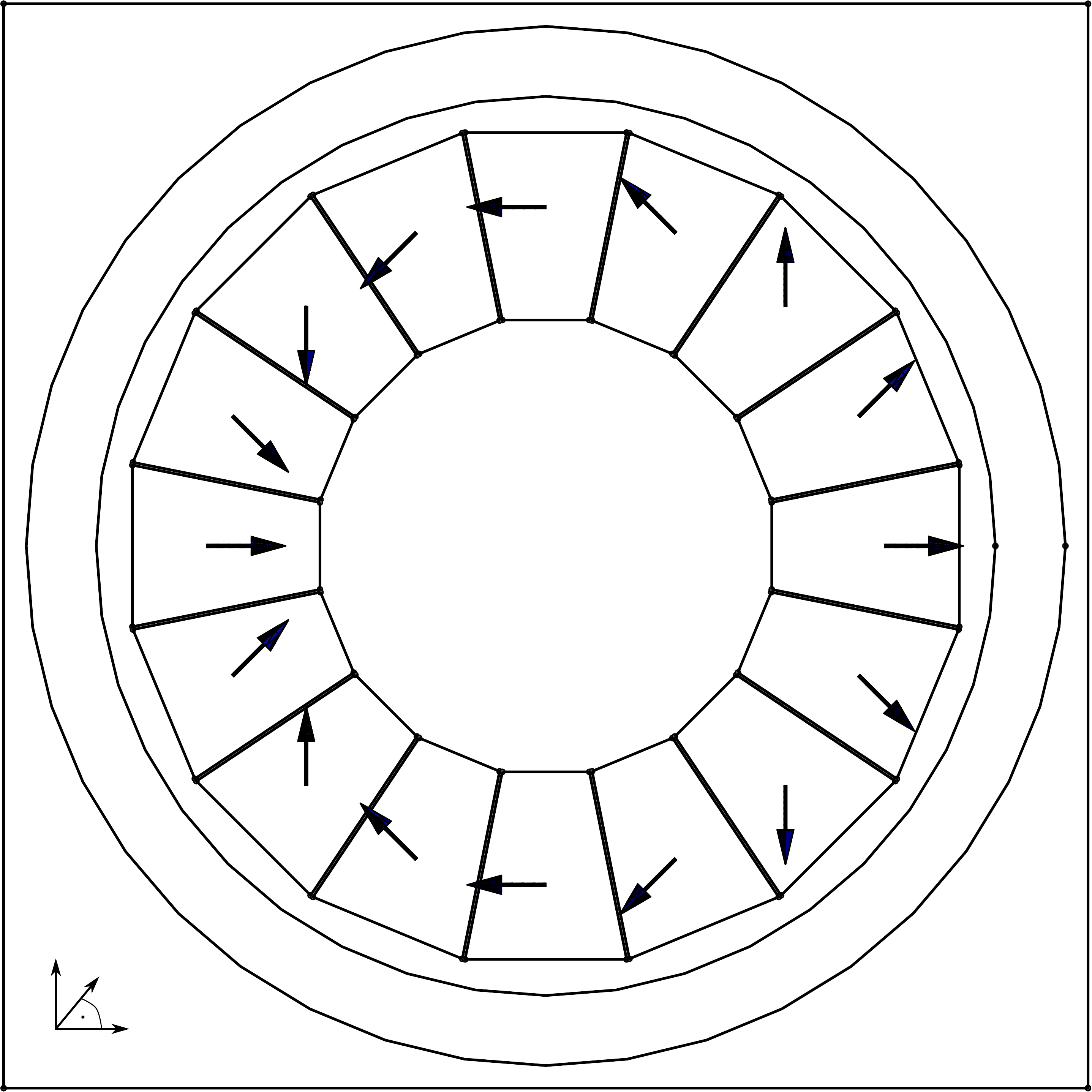}}%
    \put(0.47021672,0.49003363){\color[rgb]{0,0,0}\makebox(0,0)[lt]{\lineheight{1.25}\smash{\begin{tabular}[t]{l}$D_{\mathrm{air}}$\end{tabular}}}}%
    \put(0.45086879,0.73705804){\color[rgb]{0,0,0}\makebox(0,0)[lt]{\lineheight{1.25}\smash{\begin{tabular}[t]{l}$D_1$\end{tabular}}}}%
    \put(0.35646378,0.71747593){\color[rgb]{0,0,0}\makebox(0,0)[lt]{\lineheight{1.25}\smash{\begin{tabular}[t]{l}$D_2$\end{tabular}}}}%
    \put(0.83925346,0.86632812){\color[rgb]{0,0,0}\makebox(0,0)[lt]{\lineheight{1.25}\smash{\begin{tabular}[t]{l}$D_{\mathrm{air}}$\end{tabular}}}}%
    \put(1.7708861,0.44810127){\color[rgb]{0,0,0}\makebox(0,0)[lt]{\lineheight{1.25}\smash{\begin{tabular}[t]{l} \end{tabular}}}}%
    \put(0.4210428,0.92915605){\color[rgb]{0,0,0}\makebox(0,0)[lt]{\lineheight{1.25}\smash{\begin{tabular}[t]{l}$D_{\mathrm{iron}}$\end{tabular}}}}%
    \put(0.12784809,0.04683537){\color[rgb]{0,0,0}\makebox(0,0)[lt]{\lineheight{1.25}\smash{\begin{tabular}[t]{l}x\end{tabular}}}}%
    \put(0.02304542,0.12151899){\color[rgb]{0,0,0}\makebox(0,0)[lt]{\lineheight{1.25}\smash{\begin{tabular}[t]{l}y\end{tabular}}}}%
    \put(0.0959659,0.11261253){\color[rgb]{0,0,0}\makebox(0,0)[lt]{\lineheight{1.25}\smash{\begin{tabular}[t]{l}z\end{tabular}}}}%
  \end{picture}%
\endgroup%

%% file: Halbach_log.txt
\begingroup%
  \makeatletter%
  \providecommand\color[2][]{%
    \errmessage{(Inkscape) Color is used for the text in Inkscape, but the package 'color.sty' is not loaded}%
    \renewcommand\color[2][]{}%
  }%
  \providecommand\transparent[1]{%
    \errmessage{(Inkscape) Transparency is used (non-zero) for the text in Inkscape, but the package 'transparent.sty' is not loaded}%
    \renewcommand\transparent[1]{}%
  }%
  \providecommand\rotatebox[2]{#2}%
  \newcommand*\fsize{\dimexpr\f@size pt\relax}%
  \newcommand*\lineheight[1]{\fontsize{\fsize}{#1\fsize}\selectfont}%
  \ifx\svgwidth\undefined%
    \setlength{\unitlength}{727.51359497bp}%
    \ifx\svgscale\undefined%
      \relax%
    \else%
      \setlength{\unitlength}{\unitlength * \real{\svgscale}}%
    \fi%
  \else%
    \setlength{\unitlength}{\svgwidth}%
  \fi%
  \global\let\svgwidth\undefined%
  \global\let\svgscale\undefined%
  \makeatother%
  \begin{picture}(1.5,0.95)%
    \lineheight{1}%
    \setlength\tabcolsep{0pt}%
    \put(0,0){\includegraphics[width=\unitlength]{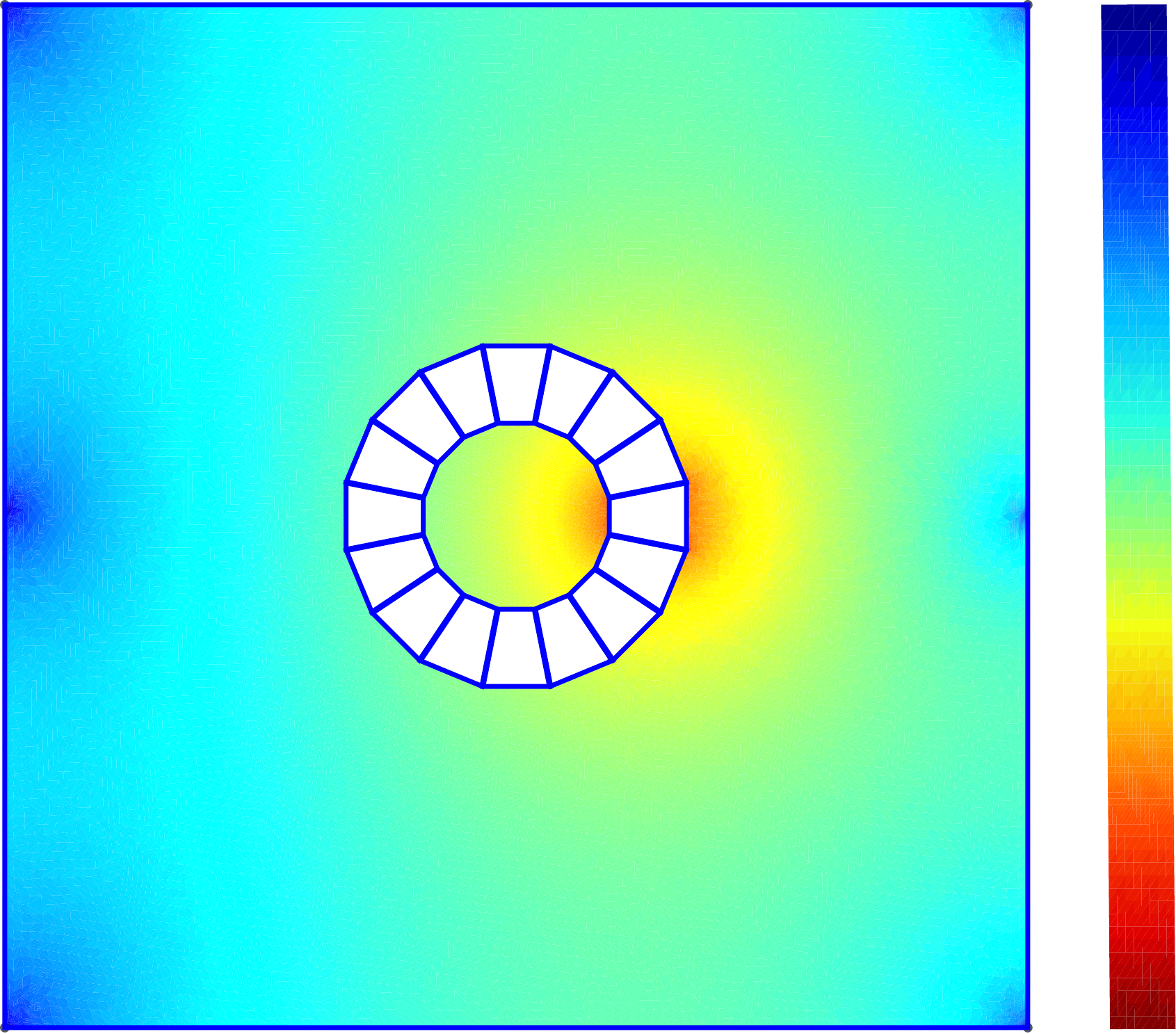}}%
    \put(1,0.79){\color[rgb]{0,0,0}\makebox(0,0)[lt]{\lineheight{1.25}\smash{\begin{tabular}[t]{l}$10^{-7}$\end{tabular}}}}%
    \put(1,0.42){\color[rgb]{0,0,0}\makebox(0,0)[lt]{\lineheight{1.25}\smash{\begin{tabular}[t]{l}$10^{-4}$\end{tabular}}}}%
    \put(1,0.02){\color[rgb]{0,0,0}\makebox(0,0)[lt]{\lineheight{1.25}\smash{\begin{tabular}[t]{l}$10^{-1}$\end{tabular}}}}%
    \put(0.9,0.92){\footnotesize\makebox(0,0)[lt]{\lineheight{1.25}\smash{\begin{tabular}[t]{l}\si{\tesla\per\A\m}\end{tabular}}}}%
  \end{picture}%
\endgroup%

%% file: lin_validation.tex
%
%
\begin{tikzpicture}

\begin{axis}[%
width=2.6in,
height=0.75in,
at={(0.758in,0.589in)},
scale only axis,
xmin=1,
xmax=16,
xlabel style={font=\color{white!15!black}},
ymin=-6000,
ymax=15000,
ylabel style={font=\color{white!15!black}},
ylabel={ $\Delta\mathbf{M}^{x}_{5,i}$ (\si{\A\per\meter})},
axis background/.style={fill=white},
axis x line*=bottom,
axis y line*=left,
xmajorgrids,
ymajorgrids,
]
\addplot [color=black, line width=1pt,,forget plot]
  table[row sep=crcr]{%
1	-2140.28991811642\\
2	-663.706294694139\\
3	-51.1987899272805\\
4	3460.04361361879\\
5	11852.4116298366\\
6	1518.27382055852\\
7	-698.537890459639\\
8	-1061.3953620988\\
9	-2059.02839064971\\
10	-744.895431885661\\
11	931.473720382043\\
12	4041.17817541256\\
13	5834.21689017254\\
14	-294.607340469927\\
15	783.496510669439\\
16	-3244.77998892221\\
};
\addplot [name path=upper,draw=none]
  table[row sep=crcr]{%
1	-3637.79489788769\\
2	-3887.25302966884\\
3	-1856.92026887091\\
4	-1680.38130926032\\
5	5774.42673758761\\
6	-2229.14510382443\\
7	-3297.76798012137\\
8	-4433.55466161706\\
9	-3716.12168208705\\
10	-4630.52812599367\\
11	-1176.94781827011\\
12	173.055738266271\\
13	2024.75830466284\\
14	-4356.538027169\\
15	-2671.96553687641\\
16	-7001.20480226616\\
};
\addplot [name path=lower,draw=none]
  table[row sep=crcr]{%
1	-642.784938345148\\
2	2559.84044028056\\
3	1754.52268901635\\
4	8600.4685364979\\
5	17930.3965220856\\
6	5265.69274494148\\
7	1900.6921992021\\
8	2310.76393741947\\
9	-401.93509921236\\
10	3140.73726222235\\
11	3039.8952590342\\
12	7909.30061255886\\
13	9643.67547568224\\
14	3767.32334622914\\
15	4238.95855821529\\
16	511.644824421749\\
};
\addplot [fill=black, fill opacity=0.2] fill between[of=upper and lower];

\addplot [color=blue, line width=1pt,,forget plot]
  table[row sep=crcr]{%
1	-2940.20292532111\\
2	1608.83668908733\\
3	8.74683568885867\\
4	6017.83356911515\\
5	13738.9606461776\\
6	3720.94180495236\\
7	-775.989307067833\\
8	-3247.28531728284\\
9	-2552.86321161196\\
10	781.398837890422\\
11	525.155656474128\\
12	9937.10998690689\\
13	3293.4577758331\\
14	2168.04201538891\\
15	900.856333626163\\
16	-2968.64619491276\\
};
\addplot [name path=upper2,draw=none]
  table[row sep=crcr]{%
1	-1648.61365858688\\
2	3117.33120320926\\
3	1319.32005145566\\
4	7207.16666083003\\
5	14854.5839697491\\
6	4982.12323053501\\
7	738.73455608952\\
8	-1640.9822315687\\
9	-1165.12034460146\\
10	2364.62835058254\\
11	1930.10455584622\\
12	11136.7965341283\\
13	4340.1686798014\\
14	3288.91000361549\\
15	2423.7751229322\\
16	-1280.93261752561\\
};
\addplot [name path=lower2,draw=none]
  table[row sep=crcr]{%
1	-4231.79219205534\\
2	100.342174965393\\
3	-1301.82638007794\\
4	4828.50047740027\\
5	12623.3373226062\\
6	2459.76037936972\\
7	-2290.71317022519\\
8	-4853.58840299697\\
9	-3940.60607862245\\
10	-801.830674801693\\
11	-879.793242897965\\
12	8737.42343968553\\
13	2246.7468718648\\
14	1047.17402716234\\
15	-622.06245567987\\
16	-4656.35977229991\\
};
\addplot [fill=blue, fill opacity=0.2] fill between[of=upper2 and lower2];

\addplot [color=green, line width=1pt,,forget plot]
  table[row sep=crcr]{%
1	-1761.21560679087\\
2	4135.82593347397\\
3	343.172329820393\\
4	5821.81708953336\\
5	13472.7733290982\\
6	5633.25159830074\\
7	882.636683533522\\
8	-1973.311666273\\
9	-1386.55547136191\\
10	292.85453264714\\
11	-612.125578594517\\
12	10696.6063034829\\
13	2996.40174303911\\
14	2717.29948193307\\
15	1608.02736241138\\
16	-4448.88301638393\\
};
\addplot [name path=upper2,draw=none]
  table[row sep=crcr]{%
1	-862.892833885586\\
2	5054.23959153681\\
3	1182.31810192079\\
4	6612.52988163258\\
5	14285.0803002506\\
6	6463.78604571299\\
7	1762.16677634711\\
8	-1027.73803815372\\
9	-447.498266030491\\
10	1222.74671632691\\
11	267.096575058349\\
12	11512.6721235729\\
13	3756.65903159062\\
14	3553.60635089322\\
15	2570.64484927796\\
16	-3487.61380580525\\
};
\addplot [name path=lower2,draw=none]
  table[row sep=crcr]{%
1	-2659.53837969616\\
2	3217.41227541113\\
3	-495.973442279999\\
4	5031.10429743413\\
5	12660.4663579458\\
6	4802.71715088849\\
7	3.10659071993473\\
8	-2918.88529439227\\
9	-2325.61267669334\\
10	-637.037651032631\\
11	-1491.34773224738\\
12	9880.54048339299\\
13	2236.1444544876\\
14	1880.99261297293\\
15	645.409875544803\\
16	-5410.15222696262\\
};
\addplot [fill=green!10] fill between[of=upper2 and lower2];

\addplot [color=red, dashed, line width=1pt, forget plot]
  table[row sep=crcr]{%
1	-2685.91665623078\\
2	3853.87303374106\\
3	292.937018019399\\
4	5145.45873662005\\
5	13942.7150132146\\
6	5246.42787245112\\
7	616.701120502319\\
8	-1324.56996502866\\
9	-1394.08950809944\\
10	1861.53455077992\\
11	242.587109528012\\
12	10644.8023078326\\
13	3588.80997030903\\
14	2087.56399156643\\
15	972.852817918191\\
16	-3808.67756021683\\
};

\end{axis}
\end{tikzpicture}%
\\
\begin{tikzpicture}

\begin{axis}[%
width=2.6in,
height=0.75in,
at={(0.758in,0.589in)},
scale only axis,
xmin=1,
xmax=16,
xlabel style={font=\color{white!15!black}},
xlabel={number of PM block $i$},
ymin=-11500,
ymax=13000,
ylabel style={font=\color{white!15!black}},
ylabel={ $\Delta\mathbf{M}^{z}_{5,i}$ (\si{\A\per\meter})},
axis background/.style={fill=white},
axis x line*=bottom,
axis y line*=left,
xmajorgrids,
ymajorgrids,
legend columns = 2,
legend style={legend cell align=left, align=left, at={(1.00,-0.4)}, draw=white!15!black, inner sep=0.3pt, style={column sep=0.01cm}}
]
\addplot [color=black, line width=1pt]
  table[row sep=crcr]{%
1	106.051294039998\\
2	-6326.50361776253\\
3	-10198.385675587\\
4	-4913.82379119064\\
5	-57.8645665232553\\
6	4253.02196568918\\
7	9349.28399407676\\
8	7615.3515775006\\
9	-88.098839536273\\
10	-7007.57675309417\\
11	-2059.93202935296\\
12	-3164.41625243226\\
13	-38.1214768217062\\
14	5185.36551756297\\
15	2010.3303595944\\
16	6718.17321478682\\
};
\addlegendentry{$\boldsymbol\mu_0-\overline{\mathbf{M}}$}
\addplot [name path=upper,draw=none,forget plot]
  table[row sep=crcr]{%
1	180.399603007978\\
2	-2747.77606415912\\
3	-3300.19513667096\\
4	-1285.39849083198\\
5	-21.0746354501351\\
6	7832.95676294633\\
7	15665.1289407275\\
8	12912.4228898466\\
9	-28.8897925854588\\
10	-1873.61574570791\\
11	4470.74148005125\\
12	222.390670085763\\
13	-2.72187358436856\\
14	8874.14548210743\\
15	11326.8351383332\\
16	10781.5215948567\\
};
\addplot [name path=lower,draw=none,forget plot]
  table[row sep=crcr]{%
1	31.7029850720187\\
2	-9905.23117136594\\
3	-17096.576214503\\
4	-8542.2490915493\\
5	-94.6544975963756\\
6	673.087168432021\\
7	3033.43904742602\\
8	2318.28026515461\\
9	-147.307886487087\\
10	-12141.5377604804\\
11	-8590.60553875717\\
12	-6551.22317495028\\
13	-73.5210800590439\\
14	1496.58555301851\\
15	-7306.17441914439\\
16	2654.82483471695\\
};
\addplot [fill=black, fill opacity=0.2] fill between[of=upper and lower];
\addlegendentry{$\boldsymbol\mu_0-\overline{\mathbf{M}}\pm \sqrt{\mathrm{diag}(\mathbf{C}_0)}$}

\addplot [color=blue, line width=1pt]
  table[row sep=crcr]{%
1	97.5635043347909\\
2	-6059.25123397233\\
3	-9175.61467160772\\
4	-5685.00412089068\\
5	-55.927477147508\\
6	1846.49636824053\\
7	8892.05580069613\\
8	9493.79543735138\\
9	-88.8714905191577\\
10	-8038.64164908053\\
11	10998.3611498252\\
12	-3731.36163849193\\
13	-41.3107305787648\\
14	6427.39531463153\\
15	3052.26295958325\\
16	1243.10148027697\\
};
\addlegendentry{$\boldsymbol\mu_1^{\mathbf{B}}-\overline{\mathbf{M}}$}
\addplot [name path=upper2,draw=none,forget plot]
  table[row sep=crcr]{%
1	171.637609604323\\
2	-4467.95685555341\\
3	-6967.63728875919\\
4	-4070.3596398794\\
5	-19.1706871451789\\
6	3431.02407709997\\
7	11039.07750232\\
8	11426.1099024729\\
9	-29.8044194114409\\
10	-6120.92879044694\\
11	13210.5055186334\\
12	-2185.41222907379\\
13	-5.94199910688481\\
14	8073.55539544229\\
15	5494.23088328609\\
16	2956.41137487304\\
};
\addplot [name path=lower2,draw=none,forget plot]
  table[row sep=crcr]{%
1	23.4893990652587\\
2	-7650.54561239126\\
3	-11383.5920544563\\
4	-7299.64860190195\\
5	-92.6842671498372\\
6	261.96865938109\\
7	6745.03409907223\\
8	7561.4809722299\\
9	-147.938561626874\\
10	-9956.35450771412\\
11	8786.216781017\\
12	-5277.31104791008\\
13	-76.6794620506448\\
14	4781.23523382078\\
15	610.295035880397\\
16	-470.208414319096\\
};
\addplot [fill=blue, fill opacity=0.2] fill between[of=upper2 and lower2];
\addlegendentry{$\boldsymbol\mu_1^{\mathbf{B}}-\overline{\mathbf{M}}\pm \sqrt{\mathrm{diag}(\mathbf{C}^{\mathbf{B}}_1)}$}

\addplot [color=green, line width=1pt]
  table[row sep=crcr]{%
1	100.922081873811\\
2	-4024.09452011583\\
3	-8571.13375957561\\
4	-3743.13978830424\\
5	-52.0206820741007\\
6	3610.33896140693\\
7	7332.33980768712\\
8	5855.27807193214\\
9	-96.1705062469244\\
10	-6924.10782483789\\
11	10874.7087896151\\
12	-3349.00683069323\\
13	-40.6240327863288\\
14	5802.0398415305\\
15	3185.67911221269\\
16	-770.530890369063\\
};
\addlegendentry{$\boldsymbol\mu_1^{\mathrm{F}}-\overline{\mathbf{M}}$}
\addplot [name path=upper2,draw=none,forget plot]
  table[row sep=crcr]{%
1	168.145905465005\\
2	-3603.9865453763\\
3	-8190.23390150657\\
4	-3319.71897580638\\
5	-16.583200824644\\
6	4027.50450583499\\
7	7694.83549623543\\
8	6281.56188910938\\
9	-41.8798744263316\\
10	-6496.03518177833\\
11	11260.0088418231\\
12	-2925.42695643416\\
13	-6.26296174008347\\
14	6234.34426183154\\
15	3599.8217636032\\
16	-348.853347423777\\
};
\addplot [name path=lower2,draw=none,forget plot]
  table[row sep=crcr]{%
1	33.6982582826164\\
2	-4444.20249485535\\
3	-8952.03361764465\\
4	-4166.56060080211\\
5	-87.4581633235574\\
6	3193.17341697886\\
7	6969.8441191388\\
8	5428.9942547549\\
9	-150.461138067517\\
10	-7352.18046789746\\
11	10489.408737407\\
12	-3772.5867049523\\
13	-74.9851038325741\\
14	5369.73542122947\\
15	2771.53646082217\\
16	-1192.20843331435\\
};
\addplot [fill=green!10,] fill between[of=upper2 and lower2];
\addlegendentry{$\boldsymbol\mu_1^{\mathrm{F}}-\overline{\mathbf{M}}\pm \sqrt{\mathrm{diag}(\mathbf{C}^{\mathrm{F}}_1)}$}

\addplot [color=red, dashed, line width=1pt]
  table[row sep=crcr]{%
1	142.467577900147\\
2	-3907.94054261331\\
3	-8761.09957959256\\
4	-3709.52912029361\\
5	-17.8025647594934\\
6	2823.95572452603\\
7	7739.24356343306\\
8	5981.63730698025\\
9	-9.48798072728263\\
10	-7260.26230830657\\
11	11370.400017288\\
12	-3800.08230561225\\
13	-33.5065060124924\\
14	6161.2003660389\\
15	2858.18666995249\\
16	-592.542403586348\\
};
\addlegendentry{$\mathbf{p}^{\mathrm{true}}-\overline{\mathbf{M}}$}

\end{axis}
\end{tikzpicture}%

%% file: update_MCMC.tex
%
%
\begin{tikzpicture}

\begin{axis}[%
width=2.8in,
height=0.75in,
at={(0.758in,0.603in)},
scale only axis,
xmin=1,
xmax=16,
xlabel style={font=\color{white!15!black}},
xlabel={number of PM block $j$},
ymin=-7100,
ymax=15950,
ylabel style={font=\color{white!15!black}},
ylabel={ $\Delta\mathbf{M}^{x}_{i}$ (\si{\A\per\meter})},
axis background/.style={fill=white},
axis x line*=bottom,
axis y line*=left,
xmajorgrids,
ymajorgrids,
legend columns = 3,
legend style={legend cell align=left, align=left, at={(1.00,-0.4)}, draw=white!15!black, inner sep=0.3pt, style={column sep=0.01cm}}
]
\addplot [color=black, line width=1.0pt]
  table[row sep=crcr]{%
1	-2568.3479017397\\
2	-796.447553632967\\
3	-61.4385479127365\\
4	4152.05233634255\\
5	14222.8939558039\\
6	1821.92858467023\\
7	-838.245468551567\\
8	-1273.67443451856\\
9	-2470.83406877965\\
10	-893.874518262793\\
11	1117.76846445845\\
12	4849.41381049508\\
13	7001.06026820705\\
14	-353.528808563913\\
15	940.195812803327\\
16	-3893.73598670665\\
17	11227.102189005\\
18	5647.37393480217\\
19	3845.61217409874\\
20	-3937.54496341147\\
21	-6855.94748604613\\
22	7265.46846425132\\
23	-5703.35484429675\\
24	-3484.67590851112\\
25	-9954.89630030034\\
26	3107.50872374284\\
27	-1125.45988950407\\
28	-5589.54744178263\\
29	-7635.84266454222\\
30	6484.33044975126\\
31	1685.13427094147\\
32	-3390.33442475967\\
};
\addlegendentry{$\boldsymbol\mu_0-\overline{\mathbf{M}}$}
\addplot [name path=upper,draw=none,forget plot]
  table[row sep=crcr]{%
1	-1306.75225603308\\
2	2719.76956311783\\
3	1916.47119612526\\
4	9521.92250529016\\
5	17480.9856785993\\
6	5859.07374173131\\
7	1988.42876569363\\
8	2383.56162322623\\
9	-961.646766755268\\
10	3346.95100410413\\
11	3381.90080854963\\
12	8595.79558169807\\
13	10041.657059049\\
14	4093.75200647835\\
15	4705.9538409698\\
16	-98.2110517859192\\
17	16343.410119801\\
18	9620.50841402478\\
19	8165.57353903426\\
20	-101.329714791097\\
21	-4448.92959770624\\
22	13054.1763559148\\
23	-1184.04112663958\\
24	652.241911442585\\
25	-8073.43978101916\\
26	5980.11205045709\\
27	5290.92358695512\\
28	-1168.09019256763\\
29	-5261.91984528701\\
30	11673.2408779143\\
31	3927.33133832614\\
32	483.382254087468\\
};
\addplot [name path=lower,draw=none,forget plot]
  table[row sep=crcr]{%
1	-3829.94354744633\\
2	-4312.66467038376\\
3	-2039.34829195073\\
4	-1217.81783260506\\
5	10964.8022330086\\
6	-2215.21657239086\\
7	-3664.91970279676\\
8	-4930.91049226335\\
9	-3980.02137080403\\
10	-5134.70004062972\\
11	-1146.36387963273\\
12	1103.03203929208\\
13	3960.46347736509\\
14	-4800.80962360618\\
15	-2825.56221536314\\
16	-7689.26092162738\\
17	6110.79425820898\\
18	1674.23945557955\\
19	-474.349190836777\\
20	-7773.76021203185\\
21	-9262.96537438602\\
22	1476.76057258787\\
23	-10222.6685619539\\
24	-7621.59372846483\\
25	-11836.3528195815\\
26	234.905397028588\\
27	-7541.84336596326\\
28	-10011.0046909976\\
29	-10009.7654837974\\
30	1295.42002158828\\
31	-557.062796443212\\
32	-7264.05110360681\\
};
\addplot [fill=black, fill opacity=0.2, forget plot] fill between[of=upper and lower];

\addplot [color=blue, line width=1pt]
  table[row sep=crcr]{%
1	-2948.52378717621\\
2	1739.2595915787\\
3	-585.493112639135\\
4	1431.53100138599\\
5	13423.1725001889\\
6	1777.19774805741\\
7	-1087.54542063469\\
8	-1805.10983119002\\
9	-2447.77388517523\\
10	562.118231512095\\
11	642.309197829674\\
12	5482.26122307062\\
13	5541.55769256755\\
14	2376.06001342048\\
15	1424.30996242895\\
16	-4332.56397206539\\
17	8354.12870529341\\
18	3926.95680927644\\
19	4892.37390845428\\
20	-2136.61783928675\\
21	-8549.92494368233\\
22	6357.86049650892\\
23	-4829.65897019136\\
24	-1525.12701880777\\
25	-11153.0795524624\\
26	1646.71099698293\\
27	-2982.21602713897\\
28	-3394.2358120762\\
29	-8211.41320048792\\
30	5787.89729400575\\
31	1773.32903784607\\
32	-3560.74766426231\\
};
\addlegendentry{$\boldsymbol\mu_1^{\mathbf{B}}-\overline{\mathbf{M}}$}
\addplot [name path=upper2,draw=none,forget plot]
  table[row sep=crcr]{%
1	-3098.38118617706\\
2	827.680213321845\\
3	-1048.46364852228\\
4	860.346845234263\\
5	12966.3008016952\\
6	1106.09074106246\\
7	-1911.26849572822\\
8	-2205.68194591304\\
9	-2726.8076889686\\
10	-131.051051820339\\
11	-45.0067820237468\\
12	5007.42880616645\\
13	4994.72840528873\\
14	1272.39971174322\\
15	848.793112702571\\
16	-4966.39041863431\\
17	7792.22767759613\\
18	3608.6563149763\\
19	4344.43053816983\\
20	-2964.18244646643\\
21	-9198.19645278368\\
22	5764.29483193127\\
23	-5673.82809524778\\
24	-2251.39220460145\\
25	-11596.0764150632\\
26	803.796239166542\\
27	-3872.71301253155\\
28	-4097.79878259607\\
29	-9187.32669501632\\
30	4660.82555093814\\
31	1087.73824841334\\
32	-3937.14516083473\\
};
\addplot [name path=lower2,draw=none,forget plot]
  table[row sep=crcr]{%
1	-2780.17087008501\\
2	2223.19881824771\\
3	58.310203187374\\
4	1947.33121592979\\
5	13963.0345728821\\
6	2679.90695406134\\
7	-440.815566560111\\
8	-1322.71290697017\\
9	-2076.81559083116\\
10	1211.18729367937\\
11	1305.71382261654\\
12	5915.39596311715\\
13	6150.19520776289\\
14	2886.66324227927\\
15	2094.28801256467\\
16	-3754.72296408205\\
17	8767.89637477938\\
18	4297.26650828359\\
19	5309.09800818987\\
20	-1595.00156312301\\
21	-7468.69355218289\\
22	6964.58713254501\\
23	-4112.99694497868\\
24	-787.052722986728\\
25	-10402.5830251642\\
26	2460.27252705925\\
27	-1945.90044225118\\
28	-2741.33786879885\\
29	-7598.2496676178\\
30	6783.54305404029\\
31	2542.70497877812\\
32	-3209.07900489619\\
};
\addplot [fill=blue, fill opacity=0.2, forget plot] fill between[of=upper2 and lower2];
\addplot [color=red, line width=1.0pt, dashed]
  table[row sep=crcr]{%
1	-2895.86507780728\\
2	864.846948160629\\
3	-661.499414750801\\
4	1768.66115453269\\
5	13120.9405572964\\
6	897.15610777433\\
7	-1469.92887739436\\
8	-1012.72299693199\\
9	-2589.67438689151\\
10	100.925816684998\\
11	346.254527885631\\
12	6048.73657782219\\
13	5165.26380336341\\
14	1204.57450905849\\
15	574.481154562492\\
16	-3640.47815913639\\
17	7204.82725714965\\
18	3628.90815833557\\
19	5876.28995969179\\
20	-823.826866359324\\
21	-7120.87084621119\\
22	8663.22942793979\\
23	-2483.34039800329\\
24	-549.684413835954\\
25	-11358.555111144\\
26	1899.7432696999\\
27	-2196.693199959\\
28	-3109.70719958095\\
29	-8934.61298168055\\
30	5877.15528667054\\
31	2824.93132383301\\
32	-3288.05383023653\\
};
\addlegendentry{$\mathbf{p}^{\mathrm{true}}-\overline{\mathbf{M}}$}
\end{axis}
\end{tikzpicture}%

%% file: update.tex
%
%
\definecolor{mycolor1}{rgb}{0.00000,0.44700,0.74100}%
\definecolor{mycolor2}{rgb}{0.85000,0.32500,0.09800}%
\definecolor{mycolor3}{rgb}{0.92900,0.69400,0.12500}%
\definecolor{mycolor4}{rgb}{0.49400,0.18400,0.55600}%
\definecolor{mycolor5}{rgb}{0.46600,0.67400,0.18800}%
\definecolor{mycolor6}{rgb}{0.30100,0.74500,0.93300}%
\definecolor{mycolor7}{rgb}{0.63500,0.07800,0.18400}%
\begin{tikzpicture}

\begin{axis}[%
width=2.8in,
height=0.8in,
at={(0.758in,0.481in)},
scale only axis,
xlabel style={font=\color{white!15!black}},
xlabel={postions on $z$-axis},
ylabel style={font=\color{white!15!black}},
xmin=0,
xmax=157,
ymode=log,
ymin=1e-05,
ymax=0.9,
ytick={0.00001,0.0001,0.001,0.01,0.1},
yminorticks=true,
xmajorgrids,
ymajorgrids,
axis background/.style={fill=white},
legend columns = 3,
legend style={legend cell align=left, align=left, at={(1.00,-0.4)}, draw=white!15!black, inner sep=0.3pt, style={column sep=0.01cm}}
]
\addplot [color=mycolor1, line width=1pt]
  table[row sep=crcr]{%
1	0.656681920002782\\
2	0.678963372223293\\
3	0.683268106350431\\
4	0.681143102794768\\
5	0.67815500054975\\
6	0.668221981417275\\
7	0.660253873865191\\
8	0.646106762398433\\
9	0.634348547023\\
10	0.61808795600575\\
11	0.599432081056893\\
12	0.584163837818936\\
13	0.563593189188936\\
14	0.546923155529586\\
15	0.525233447916943\\
16	0.501497716998388\\
17	0.484837568653367\\
18	0.460205036017894\\
19	0.443047071480489\\
20	0.41942788057964\\
21	0.395060255691331\\
22	0.381095848261817\\
23	0.357520211547867\\
24	0.347108929360756\\
25	0.326313120681492\\
26	0.319992363706278\\
27	0.300216347185525\\
28	0.276912776147627\\
29	0.264918956061664\\
30	0.220322907999366\\
31	0.170232922156593\\
32	0.0874910766816998\\
33	0.041389751115858\\
34	0.0288585353150381\\
35	0.018558753212242\\
36	0.0165606265211209\\
37	0.0118951874373846\\
38	0.0102790963758132\\
39	0.00698121487168483\\
40	0.00519150365625717\\
41	0.00487000836424166\\
42	0.00424963235853201\\
43	0.00441002665669281\\
44	0.00423618538644509\\
45	0.0042053304727143\\
46	0.00450227732005791\\
47	0.00486116772507814\\
48	0.00625596754686047\\
49	0.00754029791547322\\
50	0.0085255723017115\\
51	0.00893905086841029\\
52	0.00852231687468753\\
53	0.00749052351848604\\
54	0.0057330484251713\\
55	0.00349858072554463\\
56	0.00156769587790592\\
57	0.000349762033300049\\
58	0.000233316992503354\\
59	0.00048870876122059\\
60	0.000549922233414925\\
61	0.00037150661761516\\
62	0.000225246009232739\\
63	0.00175896379969992\\
64	0.00397799972969393\\
65	0.00618435718823506\\
66	0.00733471014915379\\
67	0.00755864401834645\\
68	0.00691193868007136\\
69	0.00569817101928368\\
70	0.00386994263361004\\
71	0.00218441935100487\\
72	0.00116119834539233\\
73	0.00114261037440986\\
74	0.00156923860456155\\
75	0.00200991600100848\\
76	0.00217501047186994\\
77	0.00214398280342733\\
78	0.00191602488982804\\
79	0.00185248557086109\\
80	0.00235860148815663\\
81	0.0031070889414709\\
82	0.00372328933946597\\
83	0.00399240496434269\\
84	0.00414157004160161\\
85	0.00420530872304386\\
86	0.00396635939183603\\
87	0.00356897531801743\\
88	0.00327090638220692\\
89	0.00344754502476041\\
90	0.00368012786514582\\
91	0.00376781165321662\\
92	0.00365393476627279\\
93	0.00325770686932298\\
94	0.00266280421079248\\
95	0.00237796422831902\\
96	0.00298365708684291\\
97	0.00436897015809814\\
98	0.00572193946266408\\
99	0.00671536311840343\\
100	0.0072538413486535\\
101	0.00737421090127134\\
102	0.00708466641758009\\
103	0.00679436492983653\\
104	0.00707198622696758\\
105	0.00790356243120619\\
106	0.00872011362137117\\
107	0.00918118867624444\\
108	0.0091020388755698\\
109	0.00826810307157742\\
110	0.00711824291513536\\
111	0.00617756430290651\\
112	0.00734691797966884\\
113	0.00978591233406311\\
114	0.0118036008138746\\
115	0.0128470463371537\\
116	0.012258383061043\\
117	0.0108718119830584\\
118	0.00729595702374946\\
119	0.00452677315476908\\
120	0.0031317971517901\\
121	0.00211310596293459\\
122	0.0022979921705921\\
123	0.00227039663207663\\
124	0.00510783318250773\\
125	0.00761596468240249\\
126	0.0160682558876284\\
127	0.00603096197028602\\
128	0.047121402429907\\
129	0.107748275120504\\
130	0.149651216390352\\
131	0.160220593194426\\
132	0.185739110762119\\
133	0.193626216638697\\
134	0.220095648329576\\
135	0.233589010810394\\
136	0.264064644983743\\
137	0.293659393363492\\
138	0.314638998657644\\
139	0.345073271696933\\
140	0.368031029858\\
141	0.398028613717429\\
142	0.42846687495657\\
143	0.450089067320864\\
144	0.478435579707037\\
145	0.499279736767009\\
146	0.524790440254272\\
147	0.548676942556903\\
148	0.566605742808285\\
149	0.588085658300576\\
150	0.603364696680776\\
151	0.62175212566815\\
152	0.633837554351814\\
153	0.646984971081973\\
154	0.658545406383993\\
155	0.66274484015251\\
156	0.665561880459869\\
};
\addlegendentry{$E^{\mathrm{rel}}(\mathbf{z},\bm\mu_0)$}

\addplot [color=mycolor2, line width=1pt]
  table[row sep=crcr]{%
1	0.660772852189954\\
2	0.682345056671112\\
3	0.686082904899306\\
4	0.683444918907131\\
5	0.679963670342683\\
6	0.669591694261951\\
7	0.661192438784525\\
8	0.64664740593976\\
9	0.634502890002812\\
10	0.617873867023907\\
11	0.598859540843042\\
12	0.583251279575847\\
13	0.562341997053631\\
14	0.545356533692091\\
15	0.523345210655046\\
16	0.499286616676658\\
17	0.482343028553316\\
18	0.457394543116857\\
19	0.439963076436985\\
20	0.416035848020331\\
21	0.391351730863047\\
22	0.377129855588889\\
23	0.353227148033293\\
24	0.342566755789708\\
25	0.321438862789691\\
26	0.314882604799199\\
27	0.29476857312389\\
28	0.271108158706851\\
29	0.258884801812976\\
30	0.213863682039126\\
31	0.163406468101425\\
32	0.0800896819114033\\
33	0.0336132495795251\\
34	0.0209554367200431\\
35	0.0105946978761754\\
36	0.00868445578331644\\
37	0.00419402480735848\\
38	0.00292267662367597\\
39	0.000143532099675202\\
40	0.000967667075316357\\
41	0.00067142259023613\\
42	0.000850950315495809\\
43	0.000424749511356463\\
44	0.000495060138743558\\
45	0.000579659444386251\\
46	0.000523154261299647\\
47	0.000649362559020276\\
48	5.93384524540413e-05\\
49	0.000770793340196608\\
50	0.00142973279118039\\
51	0.0017675459625853\\
52	0.00152358636785388\\
53	0.000960139613273113\\
54	8.17434643326051e-05\\
55	0.000708071200785342\\
56	0.000742815390606119\\
57	0.000421009865428526\\
58	7.61015338581192e-05\\
59	5.7584258710459e-05\\
60	7.07941343296491e-05\\
61	0.000404365624063839\\
62	0.000825880468527575\\
63	0.000885157061372977\\
64	0.000591817497400288\\
65	0.000250153597427837\\
66	0.000755540176394928\\
67	0.000855042184969323\\
68	0.000425127324563669\\
69	0.000333826559449634\\
70	0.00153580382043855\\
71	0.00249605092856848\\
72	0.00283957511862611\\
73	0.00243486499488183\\
74	0.00183703239461705\\
75	0.0013673307769325\\
76	0.00121547295739522\\
77	0.00122606642169886\\
78	0.00133420738664793\\
79	0.00112685418430402\\
80	0.000262071436032091\\
81	0.000648141010072691\\
82	0.00122608624486339\\
83	0.00135770899985733\\
84	0.00134702883766919\\
85	0.00128150595582791\\
86	0.000994595566037199\\
87	0.000691346313188479\\
88	0.000672328308462899\\
89	0.00126586617063738\\
90	0.00193651443234567\\
91	0.00238654571599082\\
92	0.00249103504131752\\
93	0.00209872575517496\\
94	0.00119783625439304\\
95	0.000199470521629916\\
96	0.000267389670937643\\
97	3.59805777749436e-05\\
98	0.000486144090648669\\
99	0.000770981193028339\\
100	0.000754000747904873\\
101	0.000426218852792144\\
102	0.000247734214249501\\
103	0.00090344506932514\\
104	0.000985888794144373\\
105	0.000412893756438569\\
106	0.000200981565603838\\
107	0.000444832693775364\\
108	8.18857322630187e-05\\
109	0.00115622450038253\\
110	0.00289683020256882\\
111	0.0046831676324969\\
112	0.00456632276945542\\
113	0.00315773326493023\\
114	0.00203385594870636\\
115	0.00174145578719804\\
116	0.00297543575070858\\
117	0.00491520473539488\\
118	0.00898450436277752\\
119	0.0121280839124389\\
120	0.0137410381550384\\
121	0.0148217116145045\\
122	0.0145734063991105\\
123	0.0191105000365791\\
124	0.0218783902360854\\
125	0.0243271586097909\\
126	0.0328601202868202\\
127	0.0104343118738194\\
128	0.0311451414844952\\
129	0.0926778001459687\\
130	0.135352379520129\\
131	0.146320886121515\\
132	0.172595310669092\\
133	0.181031776624901\\
134	0.2083998984184\\
135	0.222647654320283\\
136	0.254167529506219\\
137	0.284832508725382\\
138	0.306826396737515\\
139	0.338438652462955\\
140	0.362557970862339\\
141	0.393841222511874\\
142	0.425620184827415\\
143	0.448607987920848\\
144	0.478410412310123\\
145	0.500784083072975\\
146	0.527888386448004\\
147	0.553440546190765\\
148	0.573181740486\\
149	0.596504315144883\\
150	0.613839810669292\\
151	0.634288440947613\\
152	0.648731049272889\\
153	0.664286242534363\\
154	0.678383230278857\\
155	0.685603230484064\\
156	0.691533381312224\\
};
\addlegendentry{$E^{\mathrm{rel}}(\mathbf{z},\bm\mu_1^{\mathrm{F}})$}

\addplot [color=mycolor3, line width=1pt]
  table[row sep=crcr]{%
1	0.002898626\\
2	0.003246511\\
3	0.003599308\\
4	0.003999741\\
5	0.004495148\\
6	0.004994875\\
7	0.005628158\\
8	0.006270199\\
9	0.00707497\\
10	0.007927257\\
11	0.008873632\\
12	0.010063568\\
13	0.011317412\\
14	0.012894584\\
15	0.014587028\\
16	0.01650321\\
17	0.019012556\\
18	0.021654925\\
19	0.025117213\\
20	0.028931233\\
21	0.033472505\\
22	0.039639236\\
23	0.046555823\\
24	0.056305203\\
25	0.067747914\\
26	0.084409147\\
27	0.104834635\\
28	0.132279022\\
29	0.173662034\\
30	0.22372982\\
31	0.290567897\\
32	0.353365258\\
33	0.407949759\\
34	0.45280494\\
35	0.481889107\\
36	0.504234209\\
37	0.518311419\\
38	0.529413463\\
39	0.536486922\\
40	0.542174637\\
41	0.547168593\\
42	0.550933971\\
43	0.554322232\\
44	0.556883829\\
45	0.559006766\\
46	0.56089987\\
47	0.562520595\\
48	0.564521455\\
49	0.56629195\\
50	0.567774266\\
51	0.568853783\\
52	0.569413276\\
53	0.569605651\\
54	0.569393571\\
55	0.568924851\\
56	0.568593593\\
57	0.568525661\\
58	0.568640933\\
59	0.568781294\\
60	0.568890885\\
61	0.568999647\\
62	0.569193147\\
63	0.569744474\\
64	0.570569956\\
65	0.571516626\\
66	0.572029312\\
67	0.572133932\\
68	0.571827587\\
69	0.571278745\\
70	0.570477407\\
71	0.56988164\\
72	0.569743291\\
73	0.570057861\\
74	0.570473797\\
75	0.570788131\\
76	0.570867247\\
77	0.570764994\\
78	0.570469833\\
79	0.570170056\\
80	0.570134196\\
81	0.570343732\\
82	0.570609011\\
83	0.570765172\\
84	0.570910149\\
85	0.571052818\\
86	0.571075008\\
87	0.571074831\\
88	0.571185072\\
89	0.571519821\\
90	0.571818001\\
91	0.571981795\\
92	0.571990553\\
93	0.571793443\\
94	0.571415083\\
95	0.57110856\\
96	0.571187439\\
97	0.571648881\\
98	0.572084522\\
99	0.572329503\\
100	0.572333705\\
101	0.572120736\\
102	0.571702848\\
103	0.571323955\\
104	0.571292415\\
105	0.57152491\\
106	0.571623376\\
107	0.571348157\\
108	0.570558535\\
109	0.569091308\\
110	0.567148498\\
111	0.564982837\\
112	0.563702483\\
113	0.563002628\\
114	0.561916307\\
115	0.560006989\\
116	0.556756722\\
117	0.552482267\\
118	0.546228786\\
119	0.539411493\\
120	0.531948609\\
121	0.5227431\\
122	0.511387198\\
123	0.493590351\\
124	0.47065605\\
125	0.438572398\\
126	0.389955151\\
127	0.332741122\\
128	0.262617214\\
129	0.203776306\\
130	0.156326805\\
131	0.118322824\\
132	0.093386302\\
133	0.073653923\\
134	0.060478648\\
135	0.049526136\\
136	0.041939814\\
137	0.035823263\\
138	0.030461953\\
139	0.026435439\\
140	0.022811486\\
141	0.020007394\\
142	0.017654683\\
143	0.015410143\\
144	0.013675922\\
145	0.012015964\\
146	0.010702413\\
147	0.009546799\\
148	0.008442843\\
149	0.007564282\\
150	0.006710338\\
151	0.00603297\\
152	0.005367436\\
153	0.004822145\\
154	0.004350112\\
155	0.003882398\\
156	0.003504452\\
};
\addlegendentry{$\mathbf{B}^{\mathrm{meas}}_x(\mathbf{z})$ in T}

\addplot [color=mycolor4,forget plot,name path=upper1]
  table[row sep=crcr]{%
1	1\\
2	1\\
3	1\\
4	1\\
5	1\\
6	1\\
7	1\\
8	1\\
9	1\\
10	1\\
11	1\\
12	1\\
13	1\\
14	1\\
15	1\\
16	1\\
17	1\\
18	1\\
19	1\\
20	1\\
21	1\\
22	1\\
23	1\\
24	1\\
25	1\\
26	1\\
27	1\\
28	1\\
29	1\\
30	1\\
31	1\\
32	1\\
33	1\\
34	1\\
35	1\\
};

\addplot [color=mycolor5, forget plot,name path=lower1]
  table[row sep=crcr]{%
1	7e-08\\
2	7e-08\\
3	7e-08\\
4	7e-08\\
5	7e-08\\
6	7e-08\\
7	7e-08\\
8	7e-08\\
9	7e-08\\
10	7e-08\\
11	7e-08\\
12	7e-08\\
13	7e-08\\
14	7e-08\\
15	7e-08\\
16	7e-08\\
17	7e-08\\
18	7e-08\\
19	7e-08\\
20	7e-08\\
21	7e-08\\
22	7e-08\\
23	7e-08\\
24	7e-08\\
25	7e-08\\
26	7e-08\\
27	7e-08\\
28	7e-08\\
29	7e-08\\
30	7e-08\\
31	7e-08\\
32	7e-08\\
33	7e-08\\
34	7e-08\\
35	7e-08\\
};
\addplot [color=mycolor6, forget plot,name path=upper2]
  table[row sep=crcr]{%
121	1\\
122	1\\
123	1\\
124	1\\
125	1\\
126	1\\
127	1\\
128	1\\
129	1\\
130	1\\
131	1\\
132	1\\
133	1\\
134	1\\
135	1\\
136	1\\
137	1\\
138	1\\
139	1\\
140	1\\
141	1\\
142	1\\
143	1\\
144	1\\
145	1\\
146	1\\
147	1\\
148	1\\
149	1\\
150	1\\
151	1\\
152	1\\
153	1\\
154	1\\
155	1\\
156	1\\
};
\addplot [color=mycolor7, forget plot,name path=lower2]
  table[row sep=crcr]{%
121	7e-08\\
122	7e-08\\
123	7e-08\\
124	7e-08\\
125	7e-08\\
126	7e-08\\
127	7e-08\\
128	7e-08\\
129	7e-08\\
130	7e-08\\
131	7e-08\\
132	7e-08\\
133	7e-08\\
134	7e-08\\
135	7e-08\\
136	7e-08\\
137	7e-08\\
138	7e-08\\
139	7e-08\\
140	7e-08\\
141	7e-08\\
142	7e-08\\
143	7e-08\\
144	7e-08\\
145	7e-08\\
146	7e-08\\
147	7e-08\\
148	7e-08\\
149	7e-08\\
150	7e-08\\
151	7e-08\\
152	7e-08\\
153	7e-08\\
154	7e-08\\
155	7e-08\\
156	7e-08\\
};
\addplot [fill=black, fill opacity=0.2] fill between[of=upper2 and lower2];
\addplot [fill=black, fill opacity=0.2] fill between[of=upper1 and lower1];
\end{axis}
\end{tikzpicture}%